\documentclass[a4paper,11pt,tikz,border=5pt]{article}
\pdfoutput=1 

\usepackage{jcappub} 
\usepackage{slashed}                 

\usepackage{makecell}
\usepackage{aas_macros}
\usepackage{tabularx}
\usepackage[T1]{fontenc} 
\usepackage{makecell} 
\usepackage{cancel}
\usepackage{placeins}
\usepackage{mathrsfs}
\usepackage[normalem]{ulem}

\usepackage{tikz}
\usetikzlibrary{arrows,decorations.pathmorphing,backgrounds,positioning,fit,petri,automata,shadows,calendar,mindmap,
decorations.markings,calc}
\usepackage{xcolor}

\usepackage{tikz-feynman}
\tikzfeynmanset{compat=1.1.0}
\DeclareUnicodeCharacter{03B1}{\ensuremath{\alpha}}
\DeclareUnicodeCharacter{03F5}{\varepsilon}

\title{Dark Photon - ALP Freeze-in: 511 keV and H$\alpha$ Constraints}
\author[a]{Simran Arora,}

\author[b,c]{Nandini Das}
\author[b]{Sukanta Dutta,}
\author[c]{and Ashok Goyal}

\affiliation[a]{Department of Physics, University Institute of Sciences, Chandigarh University,\\ Mohali, India.}
\affiliation[b]{SGTB Khalsa College, University of Delhi, Delhi, India-110007.}
\affiliation[c]{Department of Physics and Astrophysics, University of Delhi, Delhi, India-110007.}

\emailAdd{009simranarora@gmail.com}
\emailAdd{nandinidas.rs@gmail.com}
\emailAdd{Sukanta.Dutta@sgtbkhalsa.du.ac.in}
\emailAdd{agoyal.du@gmail.com}

\abstract{We investigate a freeze-in scenario of two component dark matter consisting of an axion-like particle (ALP) and a dark photon. The dark sector connects to the Standard Model through a dimension-five ALP-dark photon interaction, while a small kinetic mixing governs dark photon decays. Solving the coupled Boltzmann equations, we determine the parameter space consistent with the observed relic abundance. We find that, for a nearly degenerate dark sector, dark photon decay into an electron-positron pair through the kinetic mixing explains the Galactic 511 keV line while the dimension five operator can source the necessary production of dark photon and ALP particles to satisfy the relic density. We further confront the model with recent H$\alpha$ observations of dwarf galaxies, together with constraints from the cosmic microwave background, diffuse gamma rays, direct detection and collider searches. We identify viable regions  of parameter space yielding $\Omega_{\rm DM}h^2\simeq0.12$, with an effective scale $\Lambda\sim10^{10}$-$10^{12}$ GeV, and dark photon lifetimes of order $10^{26}$-$10^{29}$ s, while remaining consistent with the observed 511 keV photon flux, $H\alpha$ constraints from Leo T and  all other astrophysical constraints.
}
\begin{document}
\maketitle
\flushbottom

\section{Introduction}
The existence of dark matter (DM) is firmly established through a wide range of astrophysical and cosmological observations, yet its microscopic nature remains one of the longstanding problems in modern physics. Evidence for DM is provided by galaxy rotation curves \cite{1970ApJ...159..379R}, the dynamics of galaxy clusters including the Bullet Cluster \cite{Clowe:2006eq}, the formation of large-scale structures and the anisotropies of the cosmic microwave background (CMB) \cite{Planck:2018vyg}. Collectively, these observations demonstrate the gravitational influence of DM over a wide range of length scales, while its non-gravitational interactions remain unknown. Although the Standard Model (SM) successfully describes the strong, weak and electromagnetic interactions of all known elementary particles, it does not contain a viable DM candidate, thereby providing one of the strongest motivations for physics beyond the Standard Model (BSM) (see, e.g., Refs.\,\cite{Bertone:2004pz,Bertone:2016nfn}).

\par The only information that we have about DM is the total relic density inferred from the CMB observation by Planck satellite \cite{Planck:2018vyg} given by $\Omega_{\rm DM} h^2 = 0.1200 \pm 0.001$. There are multiple possible DM candidates which have been extensively explored in the literature. Among them, Weakly Interacting Massive Particle (WIMP) \cite{Kolb:1990vq} had been the most popular choice. One of the key features of the WIMP paradigm was that, to obtain the correct relic density, the required annihilation cross section typically corresponded to models with scattering cross sections accessible to direct detection experiments.
However, the recent null results from the experiments like {\tt LUX-ZEPLIN}  \cite{4dyc-z8zf}, {\tt XENONnT} \cite{XENON:2018voc,XENON:2023cxc}, {\tt PANDA} \cite{PANDA-X:2024dlo, PhysRevLett.119.181302}, has motivated  us to explore some alternate production mechanisms.

An attractive alternative is the freeze-in mechanism \cite{Hall:2009bx,Bernal:2017kxu}, in which DM is produced non-thermally through rare decays or scatterings of particles in the thermal bath and never reaches thermal equilibrium with the SM plasma. Consequently, freeze-in DM naturally evades the stringent bounds from direct detection experiments. In many freeze-in scenarios, however, the required tiny renormalisable couplings are introduced by hand. A theoretically well-motivated alternative is provided by higher-dimensional effective interactions, where the suppression arises naturally from a heavy new-physics scale after integrating out heavy degrees of freedom. Motivated by this framework, we investigate a freeze-in scenario in which the DM abundance is generated through a dimension-five effective interaction while simultaneously addressing the Galactic $511$~keV gamma-ray excess.

\par The $511$ keV $\gamma$ ray emission from the Galactic Centre  was first  reported by a  balloon-altitude observation performed in 1970 \cite{balloon}. Subsequent measurements by CGRO/OSSE \cite{Purcell_1997} and INTEGRAL/SPI \cite{Bouchet_2010,Siegert:2019tus} also confirmed  the existence of this emission. 
Despite extensive investigations, its origin remains unresolved. Both conventional astrophysical sources and DM scenarios have been explored as possible explanations (see, for example, Refs.\,\cite{Boehm:2003bt,Leane:2022bfm,Siegert:2023wus,Aghaie:2025dgl}).

The observed $511$ keV line signal may be interpreted as to be originated  from positronium decay which can be formed from positrons and electrons. Such positrons may be produced either through DM annihilation or through the decay of unstable dark-sector particles. While annihilating DM generally provides a better description of the observed morphology, decaying DM remains a viable possibility over a well-motivated region of parameter space. In particular, dark photons decaying into electron-positron pairs through kinetic mixing with the SM hypercharge gauge boson constitute one of the most extensively studied realizations of this scenario \cite{HOLDOM1986196,Galison:1983pa}. Nevertheless, the kinetic mixing required to explain the $511$~keV signal is generally too small to account simultaneously for the observed DM abundance if the dark photon is produced through the same interaction. This motivates considering a multi-component dark sector in which DM production and the late-time origin of the Galactic positrons arise from different interactions within a unified framework.

On the other hand, recent observations have also demonstrated that diffuse H$\alpha$ emission from gas-rich dwarf galaxies provides a powerful probe of late-time DM decays. In particular, observations of the Leo~T dwarf galaxy with the Multi Unit Spectroscopic Explorer (MUSE) on the Very Large Telescope (VLT) have established stringent H$\alpha$ limits on light DM in the eV to sub-GeV mass range \cite{Leane:2025ohu}. These constraints provide an important and complementary probe of the class of models considered in this work, and are therefore incorporated in our analysis.

\par Here, we propose a minimal model framework in which an axion-like particle (ALP) and a vector gauge boson together constitute a two-component dark sector. The late-time decay of the gauge boson into an electron-positron pair provides the positrons that subsequently form positronium, thereby explaining the observed Galactic $511$ keV emission while remaining compatible with the current $H\alpha$ constraints. The DM production is governed by a non-renormalisable dimension-five portal operator, which can arise as the low-energy effective interaction. The suppression by the heavy new-physics scale leads to the freeze-in production of both dark-sector components, while a tiny kinetic mixing governs the late-time decay of the dark photon.

While there are multiple models \cite{Fayet:2004bw,Huh:2007zw,Aghaie:2025dgl,Chen:2024jbr} addressing this class of scenarios in the literature, the distinguishing feature of our framework is that a single dimension-five operator simultaneously governs the freeze-in production of both dark-sector components, whereas the tiny kinetic mixing independently controls the late-time decay of the dark photon. This minimal setup accommodates the observed dark matter relic abundance, explains the Galactic $511$ keV emission through dark-photon decay, and remains compatible with the recent $H\alpha$ constraints without introducing additional mediators or dark-sector interactions. We further take into account all relevant astrophysical, cosmological and current experimental constraints on DM.

\noindent
The article is organized as follows: In Sec.\,\ref{sec:model}, we briefly introduce the model with relevant interactions and  mention all the other observational and experimental  constraints from astrophysics, cosmology and collider data  applicable to our model. In Sec.\,\ref{sec:DM pheno}, we discuss the DM production mechanism and the corresponding freeze-in phenomenology. In Sec.\,\ref{sec:511}, we identify regions of the parameter space consistent  with the $511$ keV line explanation. In Sec.\,\ref{sec:halpha}, we discuss the compatibility of our model with $H\alpha$ constraints.  Finally we summarize our results and conclude in Sec.\,\ref{sec:sum}.

\section{The Model}
\label{sec:model}

\subsection{The Dark Lagrangian}
We consider a minimal extension of the SM by a hidden Abelian gauge symmetry $U(1)_D$ containing a massive vector gauge boson, denoted by $\gamma_D$, and ALP $a$. The dark photon acquires its mass through the St\"uckelberg mechanism \cite{Stueckelberg:1938zz,Feldman:2007wj}, while the ALP is described as a pseudo-scalar with an explicit mass term. The corresponding gauge-invariant effective Lagrangian, where the  interaction between the visible and dark sectors is mediated by the dimension-five gauge-invariant operator  $aB_{\mu\nu}\widetilde{F}_D^{\mu\nu}$, is given by
\begin{subequations}
\begin{eqnarray}
\mathcal{L}
=
\mathcal{L}_{\gamma_D}
+
\mathcal{L}_{\substack{\rm kinetic\\ \rm mixing}}
+
\mathcal{L}_{\rm ALP}
+
\mathcal{L}_{\substack{\rm Portal\\ \rm Operator}},
\label{eq:lagrangian}
\end{eqnarray}
where
\begin{eqnarray}
\mathcal{L}_{\gamma_D}
&=&
-\frac14\,F_{D \mu\nu}F_{D}^{\mu\nu}
+\frac12\, m_{\gamma_D}^{\,2} {A_D}_\mu {A_D}^\mu,
\\
\mathcal{L}_{\substack{\rm kinetic\\ \rm mixing}}
&=&
-\frac{\epsilon}{2}\,
B_{\mu\nu}F_{D}^{\mu\nu},
\\
\mathcal{L}_{\rm ALP}
&=&
\frac12(\partial_\mu a)(\partial^\mu a)
-\frac12m_a^2a^2,
\\
\mathcal{L}_{\substack{\rm Portal\\ \rm Operator}}
&=&
\frac{c_{a\gamma\gamma_D}}{4\Lambda}\,
a\,B_{\mu\nu}\Tilde{F}_{D}^{\mu\nu}.
\end{eqnarray}
\end{subequations}
Here $B_{\mu\nu}$ and ${F_D}_{\,\mu\nu}$ are the SM hypercharge and hidden dark $U(1)_D$ field tensors respectively, ${\tilde{F}_D}^{\mu\nu}$ denotes the dual of ${F_D}_{\,\mu\nu}$, and $m_{\gamma_D}$ and $m_a$ are the masses of the dark photon and ALP, respectively. $c_{a \gamma \gamma_D}$ is a dimensionless coupling and $\Lambda$ is the UV scale of our effective theory. $\epsilon$ is the dimensionless coupling which  parametrizes the kinetic mixing between $U(1)_{Y}$ and $U(1)_D$ field strength tensor. 

After electroweak symmetry breaking, the hypercharge field is related to the physical gauge fields through
\begin{equation}
B_{\mu\nu}
=
c_W F_{\mu\nu}
-
s_W Z_{\mu\nu},
\end{equation}
where $c_W=\cos\theta_W$ and $s_W=\sin\theta_W$ while $\theta_W$ is the Weinberg angle. In the phenomenologically relevant limit  $\epsilon\ll 1$,  corrections arising from kinetic mixing are negligible. Consequently, the portal operator generates both $a\gamma\gamma_D$ and $aZ\gamma_D$ interactions, which constitute the relevant vertices entering the DM production and decay processes discussed in the following sections. After diagonalizing the kinetic terms, the dark photon acquires the interaction $\epsilon e A_{D \mu} J^{\mu}_{\rm em}$, through which it couples to the electromagnetic current.

\subsection{Existing experimental and observational constraints}
\label{sec:allotherconstraints}
Before discussing the cosmological and astrophysical implications of the model, it is useful to examine whether the parameter region relevant for freeze-in DM production is already constrained by laboratory experiments and existing observations.

Dark photons with masses in the MeV-GeV range have been extensively searched for in beam-dump, fixed-target and collider experiments. Existing searches, including E137, E141, NA64, BaBar and Belle~II, together with future facilities such as LDMX, probe kinetic-mixing parameters in the approximate range
$\epsilon\sim10^{-7}$-$10^{-3}$ for
$m_{\gamma_D}\sim{\cal O}({\rm MeV}-{\rm GeV})$
\cite{Essig:2013lka,Alexander:2016aln,Banerjee:2019hmi,Belle-II:2023esi}. In the present work, however, we consider
\begin{equation}
m_{\gamma_D}\simeq 1-10~{\rm MeV},
\qquad
\epsilon\simeq10^{-24}-10^{-23},
\end{equation}
which lies many orders of magnitude below the reach of present and projected laboratory searches. Consequently, collider and fixed-target experiments do not constrain the parameter space relevant for our analysis.

Direct-detection experiments searching for nuclear recoils rapidly lose sensitivity for DM masses below approximately $1~{\rm GeV}$ because of the extremely small recoil energies. Electron-recoil searches performed by XENON1T, XENONnT, SENSEI and SuperCDMS provide the leading laboratory probes of sub-GeV DM
\cite{Aprile:2019xxb,Cui:2017nnn,Aalbers:2022fxq,Barak:2020fql}. In the present framework, however, the elastic scattering cross section is additionally suppressed by the dimension-five portal coupling,
\begin{equation}
\frac{c_{a\gamma\gamma_D}}{\Lambda}
\sim10^{-12}\ {\rm GeV}^{-1},
\end{equation}
required to reproduce the observed relic abundance through freeze-in. The corresponding scattering rates therefore remain many orders of magnitude below both current experimental limits and projected future sensitivities.

The strongest constraints on the present scenario originate from cosmological and astrophysical observations. If the dark-sector particles are not nearly degenerate in mass, the heavier state undergoes the radiative decay
\begin{equation}
\gamma_D\rightarrow a\gamma (a \rightarrow \gamma_D\gamma)
\end{equation}
 depending on the mass-hierarchy of the DM particles with a lifetime of approximately
$10^{8}$-$10^{9}$~s for MeV-scale masses and couplings compatible with freeze-in production. Such late electromagnetic energy injection modifies the ionization history of the Universe and is strongly constrained by Big-Bang nucleosynthesis and Cosmic Microwave Background observations
\cite{Slatyer:2016qyl,Poulin:2016anj}. These considerations therefore favour the nearly degenerate regime,
\begin{equation}
m_{\gamma_D}\simeq m_a,
\end{equation}
which is the scenario adopted throughout the present work.

The co-annihilation process
$a\gamma_D\rightarrow e^+e^-$
also generates secondary photons through final-state radiation. However, the corresponding  cross section is suppressed by the square of the effective  coupling and is approximately
\begin{equation}
\langle\sigma v\rangle
\sim10^{-43}\ {\rm cm^3\,s^{-1}},
\end{equation}
for the benchmark parameters ($\Lambda = 10^{12}$GeV) considered here. This cross-section is several orders of magnitude below the annihilation rate required to explain the Galactic 511~keV emission and is therefore well below the sensitivity of present indirect-detection searches.

Finally, ALP-hypercharge-dark-photon portal provides additional energy-loss channels in supernova cores through processes such as
$e^+e^-\rightarrow a\gamma_D$,
Compton scattering,
nucleon bremsstrahlung and plasmon decay. Requiring consistency with the Raffelt cooling criterion yields an approximate upper bound
\begin{equation}
\frac{c_{a\gamma\gamma_D}}{\Lambda}
\lesssim10^{-7}\ {\rm GeV}^{-1},
\qquad
m_{a},\,m_{\gamma_D}\sim0.1-100~{\rm MeV},
\end{equation}
as discussed in Ref.\,\cite{Hook:2021ous}. Since the freeze-in mechanism considered here requires
\begin{equation}
\frac{c_{a\gamma\gamma_D}}{\Lambda}
\sim10^{-12}\ {\rm GeV}^{-1},
\end{equation}
the supernova cooling bounds are readily satisfied throughout the parameter space of interest.

Having established that the parameter space relevant to the present study is consistent with existing laboratory, astrophysical, and cosmological constraints, we now investigate the cosmological evolution of the dark sector.

\section{Dark Matter phenomenology}
\label{sec:DM pheno}
\subsection{Thermalisation of the dark sector}
 In this sub-section, we discuss about the thermalisation of the dark sector. After electroweak symmetry breaking, the hypercharge portal operator generates the physical $a\gamma\gamma_D$ and $aZ\gamma_D$ interactions, giving rise to the dominant production channels
\begin{figure}[ht]
\centering

\resizebox{0.30\textwidth}{!}{
\begin{tikzpicture}
\begin{feynman}
\vertex (a);
\vertex [above left=1.8cm of a] (d){\(f\)};
\vertex [below left=1.8cm of a] (e){\(\bar f\)};
\vertex [right=2.5cm of a] (v);
\vertex [above right=1.8cm of v] (b) {\(a\)};
\vertex [below right=1.8cm of v] (c) {\(\gamma_D\)};

\diagram*{
(d) -- [fermion] (a),
(a) -- [fermion] (e),
(a) -- [boson, edge label'=\(\gamma/Z\)] (v),
(v) -- [scalar] (b),
(v) -- [boson] (c),
};
\end{feynman}
\end{tikzpicture}
}
\hfill
\resizebox{0.30\textwidth}{!}{
\begin{tikzpicture}
\begin{feynman}
\vertex (a);
\vertex [above left=1.8cm of a] (d){\(W^+\)};
\vertex [below left=1.8cm of a] (e){\(W^-\)};
\vertex [right=2.5cm of a] (v);
\vertex [above right=1.8cm of v] (b) {\(a\)};
\vertex [below right=1.8cm of v] (c) {\(\gamma_D\)};

\diagram*{
(d) -- [boson] (a),
(a) -- [boson] (e),
(a) -- [boson, edge label'=\(\gamma/Z\)] (v),
(v) -- [scalar] (b),
(v) -- [boson] (c),
};
\end{feynman}
\end{tikzpicture}
}
\hfill
\resizebox{0.25\textwidth}{!}{
\begin{tikzpicture}
\begin{feynman}
\vertex (a) {\(Z\)};
\vertex [right=2.5cm of a] (v);
\vertex [above right=1.8cm of v] (b) {\(a\)};
\vertex [below right=1.8cm of v] (c) {\(\gamma_D\)};

\diagram*{
(a) -- [boson] (v),
(v) -- [scalar] (b),
(v) -- [boson] (c),
};
\end{feynman}
\end{tikzpicture}
}

\caption{\emph{
Feynman diagrams for the freeze-in production of the dark-sector particles $a$ and $\gamma_D$ through SM scatterings and $Z$-boson decay.
}}
\label{fig:feyndiagrams}
\end{figure}
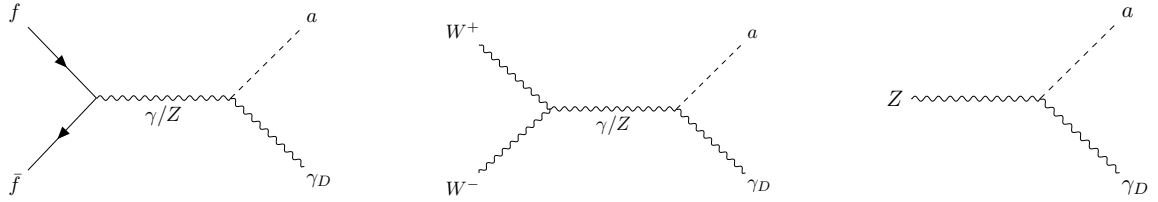
\begin{subequations}
\begin{eqnarray}
f+\bar f &\rightarrow& a+\gamma_D,\\
W^++W^- &\rightarrow& a+\gamma_D,\\
Z &\rightarrow& a+\gamma_D,
\label{eq:production_channels}
\end{eqnarray}
\end{subequations}
where $f$ denotes all SM fermions. The Feynman diagrams of the dominant processes are shown in Fig.\,\ref{fig:feyndiagrams}.  Since both dark-sector particles are assumed to have vanishing primordial abundances after reheating, these processes completely determine their subsequent cosmological evolution.

Before solving the coupled Boltzmann equations, it is preferable to verify that the benchmark parameter choices satisfy the freeze-in condition \cite{Hall:2009bx,Bernal:2017kxu}. This can be established by comparing the interaction rates of the dominant production processes with the Hubble expansion rate. During the radiation domination, the latter is given by
\begin{equation}
\mathcal H
=
\sqrt{\frac{\pi^2g_*}{90}}
\frac{T^2}{M_{\rm Pl}},
\end{equation}
where $T$ denotes the temperature of the thermal bath, $g_*$ is the effective number of relativistic degrees of freedom, and $M_{\rm Pl}=2.4\times10^{18}$ GeV is the reduced Planck mass. The interaction rates corresponding to the dominant production channels are
\begin{eqnarray}
&&\Gamma^{f\bar f}_{\rm int}
=
n_f^{\rm eq}\,\, 
\left<\sigma v\right>_{f\bar f\rightarrow a\gamma_D}; \qquad \Gamma^{W^+W^-}_{\rm int}
=
n_W^{\rm eq}\,\,
\left<\sigma v\right>_{W^+W^-\rightarrow a\gamma_D};  \nonumber\\
&&~~~~~~~~~~~~~~~~\Gamma^{Z\rightarrow a\gamma_D}_{\rm int}
=
\Gamma_{Z\rightarrow a\gamma_D}\,\,
\frac{K_1(m_Z/T)}
{K_2(m_Z/T)},
\end{eqnarray}
where $n_i^{\rm eq}$ denotes the equilibrium number density of particle $i$,
\begin{equation}
n_i^{\rm eq}
=
\frac{g_i}{2\pi^2}
m_i^2T
K_2\!\left(\frac{m_i}{T}\right),
\end{equation}
with $g_i$ being the corresponding number of internal degrees of freedom. The quantities $K_1$ and $K_2$ are the modified Bessel functions of the second kind of order 1 and order 2. For the scattering channels, the thermally averaged cross section is defined as \cite{Gondolo:1990dk}
\begin{equation}
\left<\sigma v\right>_{ij}
=
\frac{1}
{8m_i^4TK_2^2(m_i/T)}
\int_{4m_i^2}^{\infty}
ds~
\sigma_{ij}(s)
(s-4m_i^2)
\sqrt{s}
K_1\!\left(\frac{\sqrt{s}}{T}\right),
\end{equation}
where $s$ denotes the total center of mass energy of the process. The analytical expressions for all relevant production cross sections are 
\begin{eqnarray}
    \sigma_{ f \Bar{f} \to a \gamma_D} &=&
\frac{ N_c {Q^2_f} \alpha\, c_{a \gamma \gamma_D}^{\,2} c^2_{W}}
{96\,\Lambda^2}
\left(1+\frac{2m_f^2}{s}\right)
\frac{\lambda^{3/2}(s,m_a^2,m_{\gamma_D}^2)}
{s^3\sqrt{1-\frac{4m_f^2}{s}}}, \\
\sigma_{WW\to a \gamma_D}(s)
&=&
\frac{\alpha\, c_{a \gamma \gamma_D}^{\,2}c_W^2}
{288\,\Lambda^2}
\,
\frac{3s+4m_W^2}
{\sqrt{s\left(s-4m_W^2\right)}}
\,
\frac{\lambda^{3/2}(s,m_a^2,m_{\gamma_D}^2)}
{s^3}
\end{eqnarray}
where $Q_f$ and $N_c$ are the corresponding charge and the color factor of the SM fermion. Here  $\lambda$ is Kallen function defined by \(\lambda(x,y,z)=x^2+y^2+z^2-2xy-2xz-2yz\).
The decay widths used in the numerical analysis are mentioned below:
\begin{eqnarray}
    \Gamma (Z \to a \, \gamma_D) &=& \frac{c^2_{a \gamma_D \gamma} s^2_W}{384 \pi \Lambda^2 m^3_Z} \lambda^{\frac{3}{2}} (m^2_Z,m^2_{\gamma_D},m^2_a), \\
    \Gamma ( \gamma_D \to a \gamma ) &=& \frac{c^2_{a \gamma_D \gamma} c^2_W m^3_{\gamma_D}}{384 \pi \Lambda^2 } \left(1-\frac{m_a^2}{m_{\gamma_D}^2}\right)^3,
    \label{eq:gammad to a gamma} \\
    \Gamma (\gamma_D \to e^{+} e^{-}) &=& \frac{\alpha \epsilon^2}{3} m_{\gamma_D} \left(1 + \frac{2 m^2_e}{m^2_{\gamma_D}}\right)\sqrt{1- \frac{4 m^2_e}{m^2_{\gamma_D}}}
\end{eqnarray}
  
\begin{figure}[tbh]
    \centering
\includegraphics[width=1.0\linewidth]{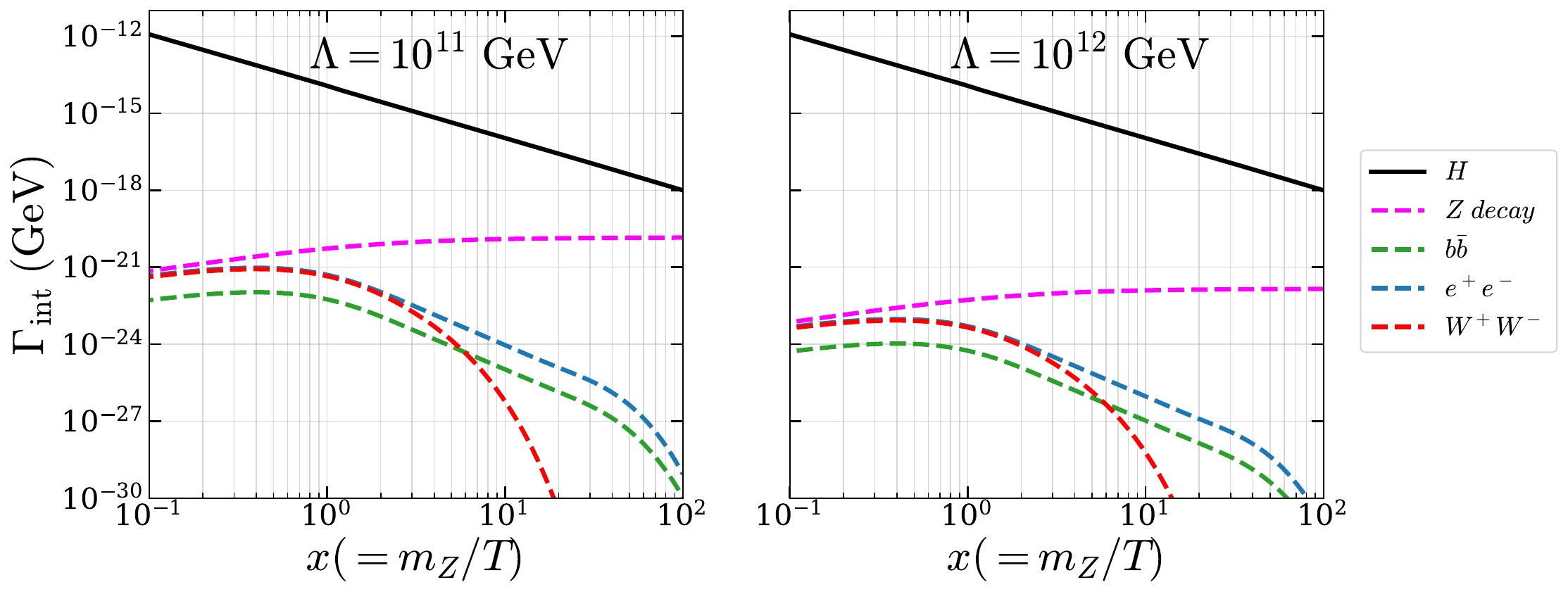}
\caption{\emph{
Evolution of the interaction rates of the dominant dark-matter production channels together with the Hubble expansion rate as functions of $x=m_Z/T$. The interaction rates remain below the Hubble expansion rate throughout the cosmological evolution, confirming that both dark-sector particles are produced via the freeze-in mechanism. The left and right panels correspond to $\Lambda=10^{11}$ GeV and $10^{12}$ GeV, respectively.}}
\label{fig:int_rate}
\end{figure}

If the interaction rate $\Gamma_{int}$ of the DM production processes are greater than the Hubble expansion rate ($\Gamma_{int} > \mathcal{H}$), it indicates that the DM would be in thermal equilibrium. In Fig.\,\ref{fig:int_rate}, we have plotted the interaction rates of the dominant DM production channels along with the Hubble expansion rate  as functions of the dimensionless variable $x=m_Z/\,T$. The two panels correspond to $\Lambda=10^{11}$ GeV and $10^{12}$ GeV respectively. The interaction rates associated with $W^+W^-$, $b\bar b$ and $e^+e^-$ annihilation decrease rapidly once the temperature drops below the masses of the corresponding initial-state particles, reflecting the Boltzmann suppression of their equilibrium number densities. In contrast, the interaction rate corresponding to $Z$-boson decay remains significantly larger than the scatterings over the whole temperature interval and therefore is expected to play a very important role in the freeze-in production of the dark sector.

For the benchmark parameter choices considered, all interaction rates remain below the Hubble expansion rate throughout the cosmological evolution, validating the freeze-in scenario. Throughout this evolution, the SM particles remain in thermal equilibrium with the radiation bath, whereas the ALP and the dark photon are populated exclusively through the feeble portal interactions and never thermalize with the SM plasma.
\subsection{Boltzmann evolution}
The cosmological evolution of the ALP and dark-photon abundances is determined by a set of coupled Boltzmann equations. The corresponding coupled Boltzmann equations in their complete form are
\begin{subequations}
\begin{eqnarray}
\frac{dn_a}{dt} 
&=& - 3\mathcal{H} n_a
+\langle \sigma v \rangle_{W^{+} W^{-}\to a\gamma_D}
\Big( (n_W^{\rm eq})^2 - n_{a} n_{\gamma_D}   \Big) + \sum_{f = \ell, q  }\langle \sigma v \rangle_{f\bar f \to a\gamma_D}
\Big( (n_f^{\rm eq})^2 - n_{a} n_{\gamma_D}   \Big) \nonumber\\
&& +\Gamma_{Z\to a\gamma_D}\,  n_Z^{\rm eq}
+ \Gamma_{\gamma_D\to a\gamma}\, n_{\gamma_D},
\label{eqn:axion eqn}
\\[2mm]
\frac{dn_{\gamma_D}}{dt} 
&=&
- 3 \mathcal{H} n_{\gamma_D}
+\langle \sigma v \rangle_{W^{+} W^{-}\to a\gamma_D}
\Big( (n_W^{\rm eq})^2 - n_{a} n_{\gamma_D}   \Big) +\sum_{f = \ell, q }\langle \sigma v \rangle_{f\bar f \to a\gamma_D}
\Big( (n_f^{\rm eq})^2 - n_a n_{\gamma_D} \Big) \nonumber\\
&&
+ \Gamma_{Z\to a\gamma_D}\, n_Z^{\rm eq}
- \Gamma_{\gamma_D\to a\gamma}\, n_{\gamma_D}
- \Gamma_{\gamma_D\to f\bar f}\, n_{\gamma_D}.
\label{eqn:vector eqn}
\end{eqnarray}
\end{subequations}
Here $n_a$ and $n_{\gamma_D}$ denote the number densities of the ALP and dark photon, respectively. In the summation, $\ell$ and $q$ denote the SM leptons and quarks. The first term on the right-hand side of each equation accounts for the dilution due to the expansion of the Universe. The next two terms describe freeze-in production from SM scatterings, while the subsequent term represents production from the decay of thermal $Z$ bosons. The remaining terms account for the decay-induced population transfer within the dark sector and the depletion of the dark photon through its decays into SM charged fermions, the latter being controlled by the kinetic-mixing parameter $\epsilon$. In the numerical analysis we employ a semi-empirical interpolation \cite{Husdal:2016haj, Drees:2015exa} for the effective relativistic degrees of freedom, incorporating both the electroweak and QCD crossover transitions at $T\simeq246$ GeV and $T\simeq160$ MeV, respectively.

Introducing the comoving yields, $Y_i=\frac{n_i}{s}$ and the dimensionless variable $x=\frac{m_Z}{T}$, the Boltzmann equations can be rewritten as
\begin{subequations}
\begin{eqnarray}\footnotesize
\frac{dY_a}{dx} \!&=& \!\! \frac{2\pi^2}{45}
\,
\frac{M_{\rm Pl}\,m_Z\,\sqrt{g_*}}
{1.66\,x^2} \Big(\langle\sigma v\rangle_{ W^+ W^-\to a \gamma_D} \big({Y_{W}^{eq}}^2 \big) +
\sum_{f = \ell, q }\langle\sigma v\rangle_{f \Bar{f}\to a \gamma_D} \big({Y_{f}^{eq}}^2 \big) \Big)
\nonumber\\
&&+ \frac{1}{\mathcal{H}x} (\langle \Gamma_{Z \to a \gamma_D} \rangle Y^{eq}_{Z} + \langle \Gamma_{\gamma_D \to a \gamma} \rangle Y_{\gamma_D} ) 
\label{eq:Ya_full}\\
\frac{dY_{\gamma_D}}{dx}\!  &=& \!\! \frac{2\pi^2}{45}
\,
\frac{M_{\rm Pl}\,m_Z\,\sqrt{g_*}}
{1.66\,x^2}
\Big(\langle\sigma v\rangle_{ W^+ W^-\to a \gamma_D} \big({Y_{W}^{eq}}^2 \big) + \sum_{f = \ell,q  }\langle\sigma v\rangle_{f \Bar{f}\to a \gamma_D} \big({Y_{f}^{eq}}^2 \big)  \Big) \nonumber\\
&&
+ \frac{1}{\mathcal{H}x} (\langle \Gamma_{Z \to a \gamma_D} \rangle Y^{eq}_{Z} - \langle\Gamma_{\gamma_D \to a \gamma} \rangle Y_{\gamma_D} - \langle\Gamma_{\gamma_D \to f \Bar{f}} \rangle Y_{\gamma_D} ). 
\label{eq:YV_full}
\end{eqnarray}
\end{subequations}
Consistent with the freeze-in hypothesis, the initial abundances of both dark-sector particles are taken to be approximately zero. Since their number densities remain several orders of magnitude below the corresponding equilibrium densities throughout the evolution, the back-reaction terms are neglected.
\begin{figure}[tbh!]
    \centering
\includegraphics[width=0.7\linewidth]{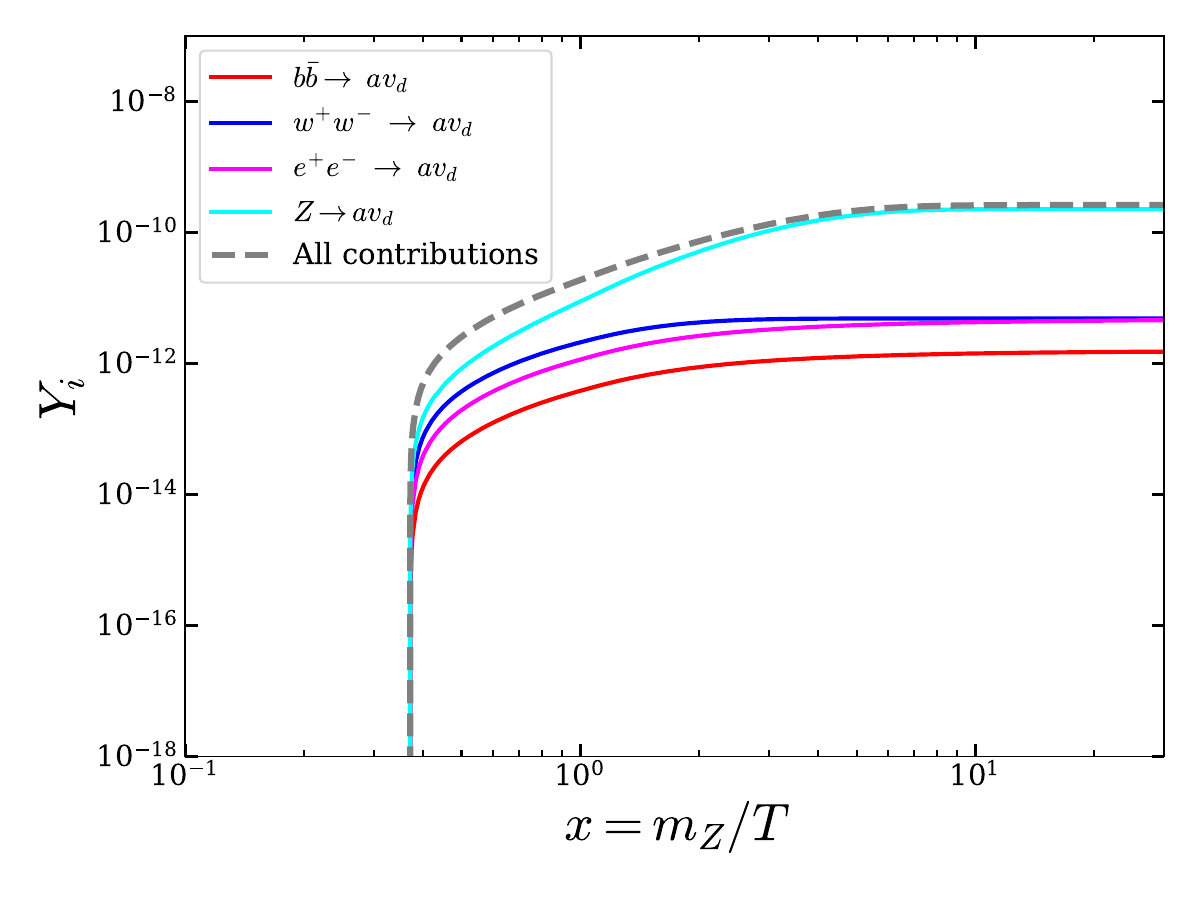}
    \caption{\emph{Evolution of the comoving yields of the dark-sector particles $a$ and $\gamma_D$ as functions of $x$. The coloured curves denote the contributions from the individual production channels, while the dashed grey curve shows the total comoving yield obtained from the coupled Boltzmann equations. The value of parameters are $\Lambda = 10^{12}$GeV and $m_a=m_{\gamma_D}=5$MeV.  }}
    \label{fig:diff_cont}
\end{figure}
The numerical solution of the coupled Boltzmann equations is presented in Fig.\,\ref{fig:diff_cont}, where the individual production channels are shown together with the total comoving abundance.

In the present work, we assume that reheating occurs above the electroweak crossover, while the freeze-in evolution is studied only after electroweak symmetry breaking, where the effective theory is expressed in terms of the physical photon and $Z$ boson. For definiteness, the numerical evolution is initiated at $T_{\rm RH}=246~\mathrm{GeV}$, corresponding to an initial value of $x\simeq0.37$. The scattering contributions saturate around $x\sim1$ owing to the Boltzmann suppression of the thermal SM particles, whereas production from thermal $Z$-boson decays continues until $x\sim10$. The decay channel $Z\rightarrow a\gamma_D$ provides the dominant contribution to the final relic abundance, consistent with the interaction-rate hierarchy shown in Fig.\,\ref{fig:int_rate}. In contrast, decay channels involving the already-produced dark-sector particles and the processes induced by kinetic mixing remain subdominant throughout the evolution. The former are suppressed by the small dark-sector abundances generated through freeze-in, while the latter are further suppressed by the tiny kinetic-mixing parameter $\epsilon$.

\begin{figure}
    \centering
    \includegraphics[width=0.6\linewidth]{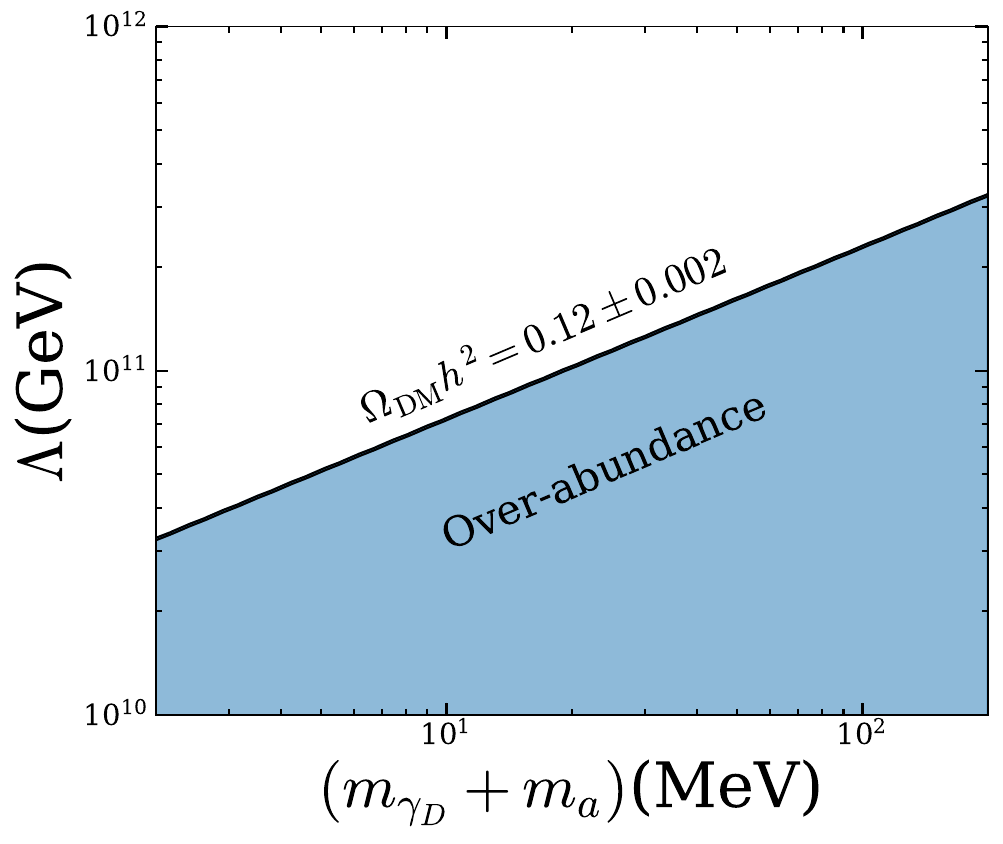}
    \caption{\emph{
Relic-density contour in the $(m_{\gamma_D}+m_a,\,\,\Lambda)$ parameter space. The black thick line corresponds to a $2 \sigma$ band around the observed dark matter abundance, $\Omega_{\rm DM}h^2=0.12$. The blue-shaded region corresponds to $\Omega_{\rm DM}h^2>0.12$ and is therefore excluded by the Planck relic-density measurement, while the region above the contour remains phenomenologically viable for the benchmark choice of model parameters.} }
    \label{fig:relic}
\end{figure}
Once the freeze-in production ceases, the comoving yields approach asymptotic values, $Y_a^\infty$ and $Y_{\gamma_D}^\infty$. The plateau in Fig.\,\ref{fig:diff_cont} marks the completion of the freeze-in process, after which the comoving yields freeze to their asymptotic values that determine the relic abundance. For the benchmark lifetimes considered here, the subsequent decay of the dark photon is negligible on cosmological timescales relevant for the relic-density calculation. The present-day dark matter relic abundance is then obtained from
\begin{equation}
\Omega_{\rm DM}h^2
=
2.742\times10^5
\left(
\frac{m_a}{\rm MeV}Y_a^\infty
+
\frac{m_{\gamma_D}}{\rm MeV}Y_{\gamma_D}^\infty
\right),
\label{eq:omega}
\end{equation}
where the numerical coefficient accounts for the present entropy density and the critical density of the Universe \cite{Bernal:2017kxu, Kolb:1990vq,Gondolo:1990dk}. The present-day dark matter abundance receives contributions from both dark-sector particles. For the benchmark parameter space considered in this work, the dark-photon lifetime is much longer than the age of the Universe, $\tau_{\gamma_D}\gg t_0$, so that its cosmological abundance remains effectively unchanged between freeze-in and the present epoch. Consequently, the relic abundance is well approximated by Eq.~(\ref{eq:omega}).

 Fig.\,\ref{fig:relic} shows the region of the $(m_{\gamma_D}+m_a,\,\Lambda)$ parameter space consistent with the observed dark matter relic abundance for the benchmark choice $c_{a\gamma\gamma_D}=1$. As expected for freeze-in production mediated by a dimension-five operator, the relic abundance decreases with increasing effective scale $\Lambda$ since the production cross sections are suppressed by $(c_{a\gamma\gamma_D}/\Lambda)^2$.  Since the freeze-in yield approximately scales as \(Y\propto \left(c/\,\Lambda\right)^2\), whereas the relic density scales as \( \Omega_{\rm DM} h^2 \propto m_{\gamma_D}\, Y\), increasing the dark-photon mass must be compensated by a  increase in \(\Lambda\), leading to the observed downward slope of the relic-density contour. The blue-shaded region corresponds to $\Omega_{\rm DM}h^2>0.12$ and is therefore excluded by the Planck measurement due to the overproduction of DM.

Having identified the parameter region consistent with the observed relic abundance, we next examine the late-time decays of the dark-sector particles, which determine the astrophysical signatures discussed in the following sections. While thermal $Z$-boson decay dominates the freeze-in production of DM, subsequent decays within the dark sector can give rise to observable cosmological and astrophysical signatures. For the mass hierarchy $m_{\gamma_D}>m_a$, the dominant dark-sector decay channel is \(\gamma_D \rightarrow a\,\gamma\), which produces a monochromatic photon with lifetime
\begin{equation}
\tau_{\gamma_D\to a\gamma}
\simeq
1.985\times10^{7}\,
c_{a\gamma\gamma_D}^{-2}
\left(\frac{\Lambda}{10^{10}\,\mathrm{GeV}}\right)^2
\left(\frac{1\,\mathrm{MeV}}{m_{\gamma_D}}\right)^3
\left(1-\frac{m_a^2}{m_{\gamma_D}^2}\right)^{-3}
\,\mathrm{s}.
\end{equation}
Such lifetimes are tightly constrained by CMB spectral-distortion
and late-time energy-injection bounds. Moreover, this decay channel cannot account for the Galactic 511 keV line, since the monochromatic photons are produced long before the present cosmological epoch and therefore do not contribute to the observed positron population in the Galactic Centre. 
We therefore do not consider this decay channel further and instead focus on the decay
 \(\gamma_D\rightarrow e^+e^-\), which is mediated by kinetic mixing. The lifetime of this decay is controlled entirely by the kinetic-mixing parameter $\epsilon$ and is approximately given by 
\begin{equation}
    \tau_{\gamma_D\rightarrow e^+e^-} \simeq 10^{28} \left( \frac{10^{-23}}{\epsilon} \right)^2 \left(\frac{2\,\mathrm{MeV}}{m_{\gamma_D}}\right)
\left(1 + \frac{2 m^2_e}{m^2_{\gamma_D}} \right)^{-1} \left(1- \frac{4 m^2_e}{m^2_{\gamma_D}}\right)^{-1/2} s
\end{equation}
For the benchmark values of $\epsilon$ considered in this work, the corresponding lifetime is of order $10^{27}\,\mathrm{s}$, consistent with the long-lived DM interpretation of the Galactic 511 keV line. In the next section we investigate whether the decay and annihilation channels of the dark sector can account for the Galactic 511 keV emission while remaining consistent with the relic-density requirement.

\section{511 keV $\gamma$-Ray Line Emission}
\label{sec:511}
The Galactic 511 keV emission line observed by the SPI spectrometer aboard the INTEGRAL satellite remains one of the most intriguing signatures of low-energy positron annihilation in the Milky Way \cite{Knodlseder:2005yq,Siegert:2015ysa}. A viable DM interpretation requires the injection of non-relativistic positrons through either DM decay or annihilation while remaining consistent with cosmological and astrophysical observations. In particular, electromagnetic energy injection before recombination is tightly constrained by BBN and CMB measurements, implying that DM decaying into photons or into a photon accompanied by a light particle must typically satisfy $\tau_{\rm DM} \gtrsim 10^{24}\text{--}10^{25}~{\rm s}$.

On the other hand, the observed 511 keV  $\gamma$-ray line can be explained through positronium decay, where the positrons are produced either via DM decay,
$
{ \rm DM} \rightarrow e^+e^-$
or DM annihilation $
{\rm DM+ \rm DM} \rightarrow e^+e^-.
$
The corresponding constraints on the decay width and annihilation cross section have been studied extensively in Ref.\,\cite{DelaTorreLuque:2023cef, Feng:2024nkh}. Requiring consistency with the observed dark matter relic abundance while explaining the 511 keV signal through positronium decay typically leads to the following annihilation cross section:
\begin{equation}
(\sigma v)_{{ \rm DM + \rm DM}\rightarrow e^+e^-}
\simeq
10^{-30}
\left(\frac{m_{\rm DM}}{\rm MeV}\right)^2
{\rm cm}^3{\rm s}^{-1}.
    \end{equation}
Similarly, if the 511 keV photon flux originates from DM decay, the corresponding lifetime is required to satisfy
\begin{equation}
\tau_{{\rm DM}\rightarrow e^+e^-}
\simeq
\left(\frac{\rm MeV}{m_{\rm DM}}\right)
10^{28}~{\rm s}.
\end{equation}

In the context of the model considered here, the electron-positron pair production can arise from the decay of vector DM $\gamma_D$ through kinetic mixing, as well as from the co-annihilation of ALP and dark photon.  In our case, the velocity averaged co-annihilation cross-section of the DM candidates to a pair of electron and positron in non-relativistic limit at \(s\approx (m_a+m_{\gamma_D})^2\) is given by
\begin{align}
\sigma v_{\rm NR}
&=
\frac{4}{3\Lambda^2}\,\alpha Q_f^2 c_{a\gamma\gamma_D}^2\sqrt{\Delta}
\Bigg[
\Delta\left(
1+\frac{m_a m_{\gamma_D}}
{(m_a+m_{\gamma_D})^2}
\right)
\nonumber\\
&\qquad
+\frac{m_e^2}{4m_a m_{\gamma_D}}
\lambda\!\left(
1,
\frac{m_a^2}{(m_a+m_{\gamma_D})^2},
\frac{m_{\gamma_D}^2}{(m_a+m_{\gamma_D})^2}
\right)
\Bigg]
\nonumber\\
&\approx
1.14\times10^{-43}\,
c_{a\gamma\gamma_D}^2
\left(\frac{10^{12}\,\mathrm{GeV}}{\Lambda}\right)^2
\,\mathrm{cm^3\,s^{-1}}.
\end{align}
where 
\begin{equation}
    \Delta = (1- \frac{4 m_e^2}{s}). \nonumber
\end{equation}

For the benchmark parameters considered in this work, the resulting cross section is  several orders of magnitude below the value required to account for the observed Galactic 511 keV flux. We are therefore led to examine the decay option carefully. In the model under consideration, we have two DM candidates with three possible decay scenarios depending on their mass hierarchy; namely   
\begin{itemize}
    \item $m_{\gamma_D}> m_a$ :
    In this case, $\gamma_D$ would decay to a photon and an ALP and the corresponding decay width is mentioned in \ref{eq:gammad to a gamma}. The coupling required to have the correct relic density by freeze-in production would result into a life time of $10^8-10^9$s for a MeV scale $\gamma_D$ with mass difference of MeV scale. However, this scenario is highly restricted by CMB observation and therefore cannot  explain the 511 keV line.
    \item $m_{a} > m_{\gamma_D}$ :
    This possibility similarly results in a significantly small lifetime of the DM candidate therefore disturbing the observation of CMB.
    \item $m_{a}\simeq m_{\gamma_D}$:
    Although this scenario requires a mild mass degeneracy between the ALP and the dark photon, it kinematically suppresses the radiative decays within the dark sector while allowing the decay $\gamma_D\rightarrow e^+e^-$ through kinetic mixing. For suitable values of $m_{\gamma_D}$ and $\epsilon$, positrons are produced through both the decay of $\gamma_D$ and the co-annihilation process $a\gamma_D\rightarrow e^+e^-$.  Once injected into the interstellar medium, $[93-97]\%$ of these positrons undergo thermalization in the Galactic bulge before binding into para-positronium, which subsequently decays into two mono-chromatic 511~keV photons. 

\end{itemize}
Since the dominant freeze-in processes produce one ALP and one dark photon per interaction, the two dark-sector particles are populated simultaneously through identical source terms in the coupled Boltzmann equations. Furthermore, as the ALP and dark photon are nearly degenerate in mass and the decay of the dark photon is negligible during the freeze-in epoch, their asymptotic relic abundances are expected to be approximately equal.
For the benchmark parameter space considered in this work, the numerical solution of the coupled Boltzmann equations yields nearly identical asymptotic comoving abundances for the ALP and the dark photon. Together with the assumed near mass degeneracy, this implies
\begin{equation}
f_a \simeq f_{\gamma_D} \simeq 0.5,
\end{equation}
which is adopted throughout the subsequent phenomenological analysis. Here $f_i$ denotes the fractional contribution of the $i^{\rm th}$ dark-sector component to the total dark matter relic abundance.

\begin{figure}
    \centering
    $$
    \includegraphics[width=0.5\linewidth]{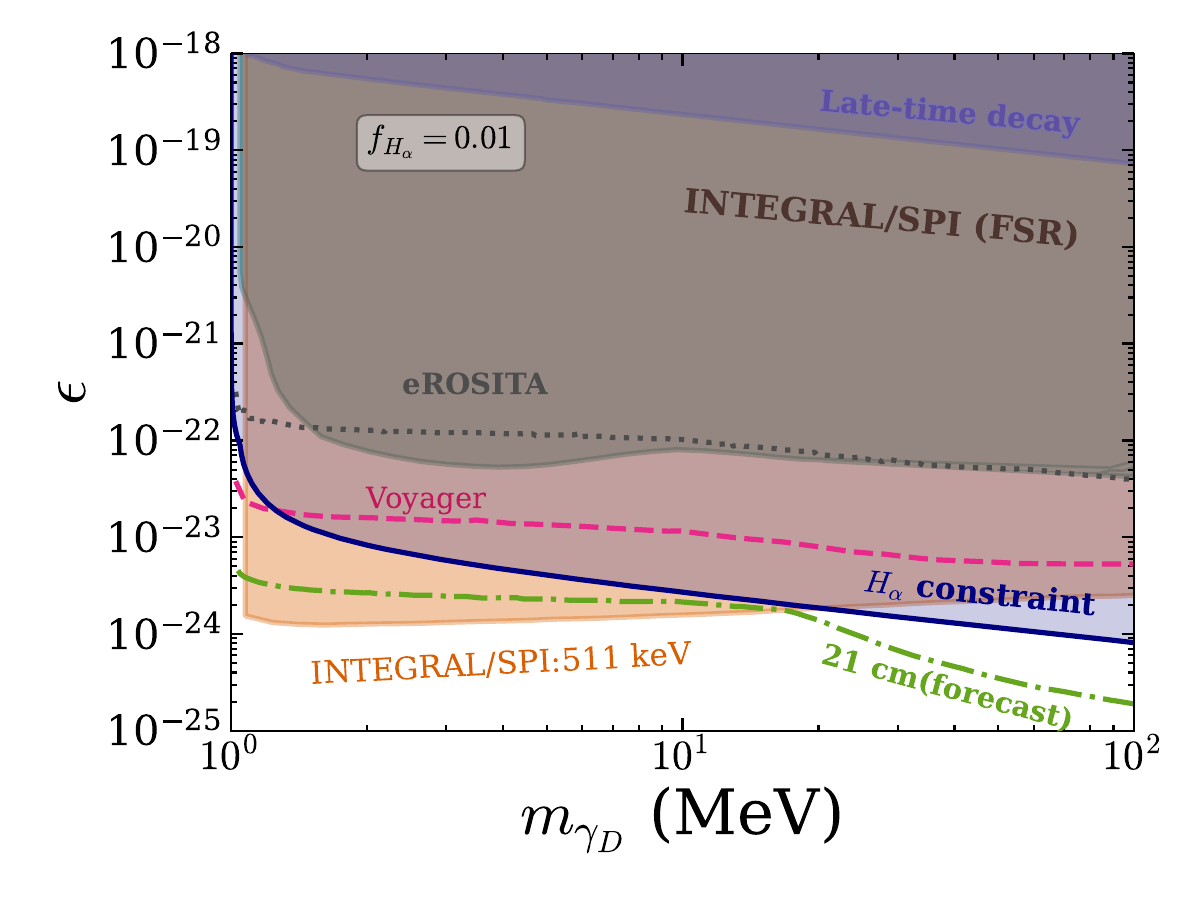}~~
    \includegraphics[width=0.5\linewidth]{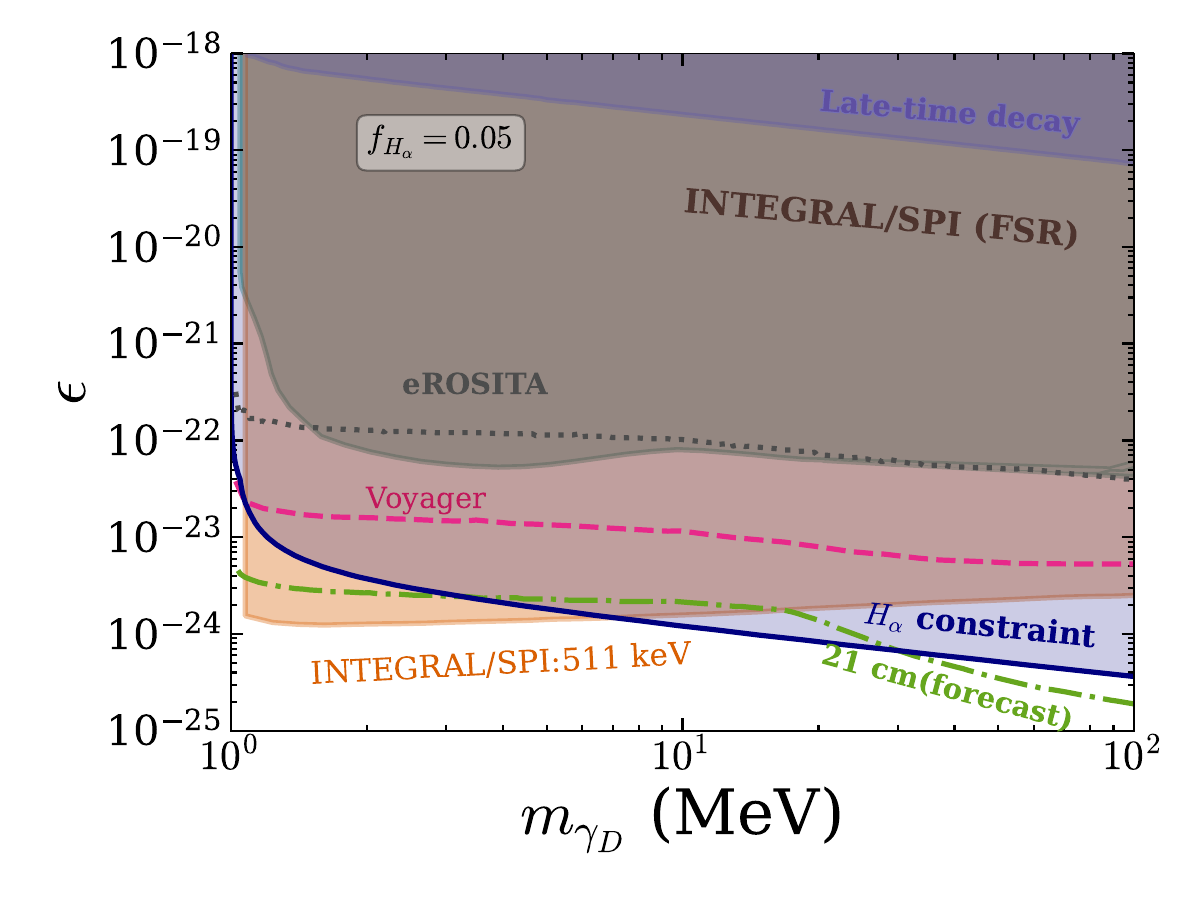}
    $$
    \caption{\emph{Combined constraints from the Galactic 511~keV emission and the $H\alpha$ observations along with all other possible competitive  constraints in the $(m_{\gamma_D},\,\epsilon)$ plane. In the left panel, the value of $f_{H\alpha}$ is 0.01 whereas $f_{H\alpha}=0.05$ in the right panel. The orange contour under the blue curve in both panels correspond to the region of parameter space where the $511$ KeV excess can be explained from the decay of dark photon while it still remains allowed from the $H\alpha$ constraints. The colored shaded regions denote the exclusion regions corresponding to experimental observations. The experimental bounds are taken from Ref.\,\cite{Nguyen:2025tkl}.    
}}
 \label{fig:511keV}
\end{figure}

\section{$H\alpha$ Emission from Dwarf Galaxies}
\label{sec:halpha}
$H\alpha$ line  originates from the $n = 3$ to $n = 2$ transition in  atomic hydrogen as emission signal in gas rich dwarf galaxies with an energy 1.89 eV . This is the brightest light that imparts red glow to the nebulae. In the dwarf galaxies the $n =2$ population is negligible and the $H\alpha$ signal can travel at large distances and escape unattenuated preserving its morphology. In the present model, the dominant source of ionizing electrons arises from the decay
$\gamma_D\rightarrow e^+e^-$,
which is mediated by the kinetic-mixing interaction. The injected electrons and positrons ionize the surrounding neutral hydrogen, ultimately leading to H$\alpha$ emission through radiative recombination. 

Among nearby dwarf galaxies, Leo~T provides one of the most sensitive environments for probing MeV-scale DM owing to its large neutral-hydrogen content and extremely low astrophysical background \cite{Ryan-Weber:2007guk}. Recent observations with the MUSE mounted on the VLT have placed stringent upper limits on diffuse H$\alpha$  flux as low as $7.8 \times 10^{-18}$ erg $\mathrm{s}^{-1} \mathrm{cm}^{-2}$, from dwarf galaxies, thereby constraining dark-matter-induced ionization \cite{Blum:2022dxi,Leane:2025ohu}, giving the constraint on the flux $\Phi_{H\alpha}$ as
$\Phi_{H\alpha}^{\rm DM}\leq \Phi_{H\alpha}^{\rm Observed}.$

For the parameter region of interest, the dominant contribution to the H$\alpha$ signal arises from
$\gamma_D\rightarrow e^+e^-$,
corresponding to dark-photon lifetimes in the approximate range \( 10^{26}\ {\rm s}\lesssim
\tau_{\gamma_D}
\lesssim10^{29}\ {\rm s}\).

The total $H\alpha$ flux from Leo T, for example, dominated by the decay induced contribution is given by \cite{Leane:2025ohu}
\begin{subequations}
\begin{equation}
    \Gamma_{H\alpha} \sim \Gamma_{H\alpha^{decay}} = \frac{f_{H\alpha} f_{\rm eq} \epsilon_{\rm dec}}{4\pi}\frac{D}{m_{\gamma_D}\tau_{\gamma_D}},
\end{equation}
where
\begin{equation}
    D = \int_{\Delta \Omega} d\Omega \int_{loc}dl \quad \rho_{\rm DM}(l, \Omega)
\end{equation}
\end{subequations}
Here $\epsilon_{\rm dec}$ is the SM injected energy to the particles which is $m_{\gamma_D} - 2 m_e$. $D$ factor  for Leo T dwarf galaxy is adopted from the benchmark astrophysical factor  as $D = 5.01 \times 10^{16}$ GeV cm$^{-2}$ \cite{PhysRevD.93.103512}. As only half of the fraction of the total DM contributes in the $H\alpha$ flux, we have multiplied the flux with a half factor. The $f_{eq}$ factor encapsulates the fact if the DM-powered ionization and the hydrogen recombination are in equilibrium or not. $f_{H\alpha}$ is the fraction of SM energy that goes into $H\alpha$. We have taken $f_{eq} \simeq 1$ and considered some reasonable  conservative benchmark values for $f_{H\alpha}$.

For the benchmark value $f_{H\alpha}=0.01$, the orange 511~keV contour in the left panel of Fig.\,\ref{fig:511keV} remains partially consistent  with the H$\alpha$ constraint, allowing dark-photon masses up to approximately $10$ MeV. Larger values of $f_{H\alpha}$ strengthen the H$\alpha$ bound and progressively shrink the overlap between the two allowed regions as shown in the right panel of Fig.\,\ref{fig:511keV} for $f_{H\alpha}=0.05$. The other experimental constraints are taken from Ref.\,\cite{Nguyen:2025tkl}. Here we would like to mention that the parameter space shown here is safe from  all the existing constraints discussed in Subsection.~\ref{sec:allotherconstraints}.

\section{Summary}
\label{sec:sum}

We have investigated a minimal two-component freeze-in dark matter framework consisting of an ALP and a dark photon interacting with the SM through a dimension-five ALP-hypercharge-dark-photon  portal, while a tiny kinetic mixing governs the late-time decay of the dark photon. The distinguishing feature of this framework is that it simultaneously produces both DM components via freeze-in, whereas the suppressed kinetic mixing independently controls the observable astrophysical signatures. This economical setup provides a unified description of the dark matter relic abundance, the Galactic $511~\mathrm{keV}$ emission, and the recently proposed H$\alpha$ constraints without introducing additional dark-sector interactions or mediators.

We solved the coupled Boltzmann equations governing the cosmological evolution of the ALP and dark-photon abundances and demonstrated that the dark sector never thermalises with the SM plasma for our choice of parameter space. As shown in Fig.\,\ref{fig:int_rate}, the interaction rates of all dominant production channels remain below the Hubble expansion rate throughout the relevant temperature range, confirming the consistency of the freeze-in scenario. The numerical evolution of the comoving yields, presented in Fig.\,\ref{fig:diff_cont}, shows that thermal $Z$-boson decay dominates the production of the dark sector, while the contributions from SM scattering processes remains subdominant after the electroweak epoch. The observed relic abundance is reproduced for MeV-scale dark-sector masses with an effective scale $\Lambda \sim 10^{10}$--$10^{12}~\mathrm{GeV}$, as illustrated in Fig.\,\ref{fig:relic}.

The Galactic signal in this model is obtained by the decay $\gamma_D\rightarrow e^+e^-$ mediated through kinetic mixing. We find that a nearly degenerate ALP--dark-photon spectrum kinematically suppresses rapid radiative decays within the dark sector while allowing dark-photon lifetimes of order $10^{26}$--$10^{29}~\mathrm{s}$, consistent with the lifetime required to explain the observed Galactic $511~\mathrm{keV}$ emission.

The recent H$\alpha$ observations of gas-rich dwarf galaxies provide a complementary probe of long-lived MeV-scale DM. Combining these constraints with those from the Galactic $511~\mathrm{keV}$ line, the cosmic microwave background, diffuse $\gamma$-ray observations, direct-detection experiments, collider searches, and supernova cooling, we identify viable regions of parameter space where all current experimental and observational constraints are simultaneously satisfied. As shown in Fig.\,\ref{fig:511keV}, an overlap remains between the parameter space favoured by the Galactic $511~\mathrm{keV}$ signal and that allowed by the H$\alpha$ observations for conservative values of the H$\alpha$ efficiency factor, corresponding to dark-photon masses up to approximately $\mathcal{O}(10)~\mathrm{MeV}$.

Our analysis demonstrates that a remarkably economical dark sector, characterised by a single dimension-five ALP- dark photon anomalous portal together with a tiny kinetic mixing, can simultaneously account for freeze-in DM production and observable late-time astrophysical signatures. The interplay between cosmological relic-density measurements and indirect probes such as the Galactic $511~\mathrm{keV}$ emission and H$\alpha$ observations provides a powerful strategy for testing this class of freeze-in dark matter models. Future improvements in MeV-scale indirect searches, together with increasingly sensitive H$\alpha$ observations of nearby dwarf galaxies, will further probe the viable parameter space identified in this work.

\section*{Acknowledgement}
ND would like to thank Debajyoti Choudhury for fruitful discussions and feedback. We acknowledge partial financial support from the ANRF grant CRG/2023/008234.  SD would like to acknowledge the University of Delhi project grant IoE/2025-26/12/FRP.
\bibliographystyle{JHEP}
\bibliography{ref}
\end{document}